\documentclass[aps,prl,twocolumn,preprintnumbers]{revtex4-2}
\usepackage{amsmath, mathrsfs, amssymb,amsfonts,amsthm,graphicx, epsf, dcolumn, yfonts}
\usepackage[hyperfootnotes=true]{hyperref}
\usepackage{color}
\usepackage[usenames,dvipsnames]{xcolor}
\hypersetup{
  colorlinks   = true, 
  urlcolor     = blue, 
  linkcolor    = blue, 
  citecolor   = red 
}
\usepackage{slashed}
\usepackage{setspace}
\usepackage{cancel}
\usepackage{wasysym}
\usepackage[lofdepth,lotdepth]{subfig}
\usepackage{float}
\usepackage{ragged2e}
\DeclareCaptionJustification{justified}{\justifying}
\pdfoutput=1
 \newcommand{\be}{\begin{equation}}
 \newcommand{\ee}{\end{equation}}
 \newcommand{\bea}{\begin{eqnarray}}
 \newcommand{\eea}{\end{eqnarray}}

\newcommand{\beq}{\begin{equation}}
\newcommand{\eeq}{\end{equation}}

\def\l{\lambda}

\def\RV{R_{{V}}}

\renewcommand*{\thefootnote}{\fnsymbol{footnote}}

\begin{document}

\title{Partition function for    a volume of space}

\author{Ted Jacobson$^{1,2}$  and Manus R. Visser$^2$}
\affiliation{$^1$Maryland Center for Fundamental Physics,   University of Maryland, College Park, MD 20742, USA \\
$^2$Department of Applied Mathematics and Theoretical Physics, University of Cambridge,
Wilberforce Road, Cambridge CB3 0WA, UK}

\begin{abstract}\vspace{-2mm}
\noindent We consider the quantum gravity partition function that counts the dimension of 
the Hilbert space of a spatial region with topology of a ball and fixed proper volume,
and evaluate it in the leading order saddle point approximation. The result is the exponential of the Bekenstein-Hawking entropy associated with 
the area of the saddle ball boundary, and is reliable within effective
field theory provided the mild curvature singularity at the ball boundary is regulated
by higher curvature terms.
This generalizes 
the classic Gibbons-Hawking computation of the de Sitter entropy for
the case of positive cosmological constant and unconstrained volume, 
and hence exhibits the holographic nature of  nonperturbative 
quantum gravity in generic finite volumes of space.

\end{abstract}

\renewcommand*{\thefootnote}{\arabic{footnote}}
\setcounter{footnote}{0}

\maketitle

\noindent \emph{Introduction.}
Gibbons and Hawking (GH) \cite{Gibbons:1976ue} proposed 
in 1977 that 
the thermal partition function $Z$ in
quantum gravity can be approximated 
by a Euclidean saddle point of a path integral over spacetime geometries.
When applied to spacetimes without a boundary, 
using the Einstein-Hilbert action $I$ with a positive cosmological constant $\Lambda$, they
found that 
there is a saddle corresponding to a round Euclidean 4-sphere, which can be obtained by 
analytic continuation of the time coordinate in a static patch of Lorentzian de Sitter (dS) spacetime. This
 method yields 
 \beq\label{Zsim}
 Z\approx \exp(-I_{\rm saddle}/\hbar)=\exp(A/4 \hbar G),
 \eeq
where $A$ is the area  of the 
dS horizon for the saddle  and $G$ is the gravitational constant  (in units with 
the speed of light equal to one).  Each ``time'' slice of the Euclidean saddle is 
a spatial geometry that is identical to the spatial geometry of a static
slice bounded by the event horizon of the corresponding Lorentzian geometry.
In the Euclidean solution 
these timeslices all coincide at this event horizon surface, which we refer to as
the ``Euclidean horizon,'' or just ``horizon''.

This and other results  
 strongly suggested
that the  concept of Bekenstein-Hawking entropy $A/4 \hbar G$ for black hole horizons \cite{Bekenstein:1973ur,Hawking:1975vcx} applies also to de Sitter horizons~\cite{Gibbons:1977mu}. 
Two decades later  Fischler \cite{Fischler}  and  Banks~\cite{Banks:2000fe} argued that, since the sphere partition function is 
an integral unconstrained by any boundary conditions,
this $Z$ must represent the dimension of the Hilbert space
of {\it all} states in this theory describing  a volume of space, whose size in the saddle point approximation is determined by the value of the cosmological constant.  That is,   $\log Z$ is the
entropy of the maximally mixed state in that Hilbert space. 
This interpretation of the ``entropy of de Sitter space"
has since received support from several directions \cite{Banks:2003ta,Banks:2006rx,Dong:2018cuv,Banihashemi:2022jys,Chandrasekaran:2022cip,Lin:2022nss}.

Numerous lines of evidence indicate  that  gravitational entropy is associated not only to  the area of a de Sitter or black hole horizon, but also to the area of any boundary separating a  region of space (see for example~\cite{Jacobson:2003wv,Bianchi:2012ev}),
in particular to the bounding area of a topological ball of space.  We present a  derivation of the 
``entropy of a volume of space'' from a quantum gravity  partition function, without or with a cosmological constant, in the vein of the original GH calculation. In our case, the size of the spatial region is determined not by 
a cosmological constant, but by a volume constraint imposed on the states. The constraint modifies the saddle point condition, and we find that the Einstein-Hilbert action of the  
saddle geometry is $-A/4  G$, where $A$ is the boundary area.
Our derivation involves no classical background spacetime as input; rather, it is in principle a fully nonperturbative quantum gravity calculation from which a classical saddle arises as output. On the other hand, like the GH calculation, it is formulated within general relativity, which is  the 
low-energy effective theory of some ultraviolet-complete theory of quantum gravity.  

The microstates counted by the Bekenstein-Hawking entropy are arguably related to vacuum fluctuations \cite{Sorkin:1984kjy,Bombelli:1986rw,Frolov:1993ym,Jacobson:2012ek,
Cooperman:2013iqr},  
but their precise nature is not resolvable within the low-energy effective theory. 
Moreover, the notion of a Hilbert space of a subregion in quantum gravity requires explanation. 
Because of the diffeomorphism constraints, the Hilbert space of quantum gravity is not decomposable as a tensor product of subregions of Hilbert spaces. This is similar to the case of Yang-Mills theory, for which the entire Hilbert space can nevertheless be realized as   a gauge-invariant \emph{subspace} of a tensor product of subregion Hilbert spaces that contain ``edge state'' degrees of freedom \cite{Donnelly:2011hn,Donnelly:2014gva}.  We presume that the Hilbert space of a ball of space in 
quantum gravity admits (at least in some approximate description) 
a similar subregion interpretation, 
and that horizon entropy can be thought of as the logarithm of the dimension of the space of edge states. The remarkable thing is that, even without having the
microscopic theory of the edge states, their contribution to the entropy per unit horizon area may be encoded in the 
value of the low-energy effective gravitational constant~\cite{Susskind:1994sm,Jacobson:1994iw}.

\newpage
 \noindent \emph{Sphere partition function.}   The GH partition function  
for the case of dS---which has no boundary---can be interpreted as 
the trace of the identity operator on the Hilbert space of {\it all} states
of a topological ball. This interpretation 
has recently been justified~\cite{Banihashemi:2022jys}
by first introducing an artificial internal boundary sphere of radius $R_B$, and
inverse temperature $\beta$ at that boundary, and
considering the canonical partition function $Z= {\rm Tr} \exp(-\beta H_{\rm BY})$,
where $H_{\rm BY}$ is the Brown-York Hamiltonian for that system.
In the limit $R_B\rightarrow0$ the boundary disappears and 
$H_{\rm BY}\rightarrow0$, 
assuming the geometry is regular inside the shrinking boundary
and assuming $D\ge 3$ spacetime dimensions.
Therefore, in this limit, $Z \rightarrow {\rm Tr}\exp(0)={\rm Tr}\, I_{\mathcal H}$, 
where  ${\mathcal H}$ is the Hilbert space of the ball of space.
That is, the  no-boundary canonical ensemble is maximally mixed, 
and $Z$ counts the dimension of the entire Hilbert space.

\begin{figure}[t!]
 \begin{center}
 \includegraphics[width=.2\textwidth, trim = 4.5cm 3cm 3.5cm 3.5cm]{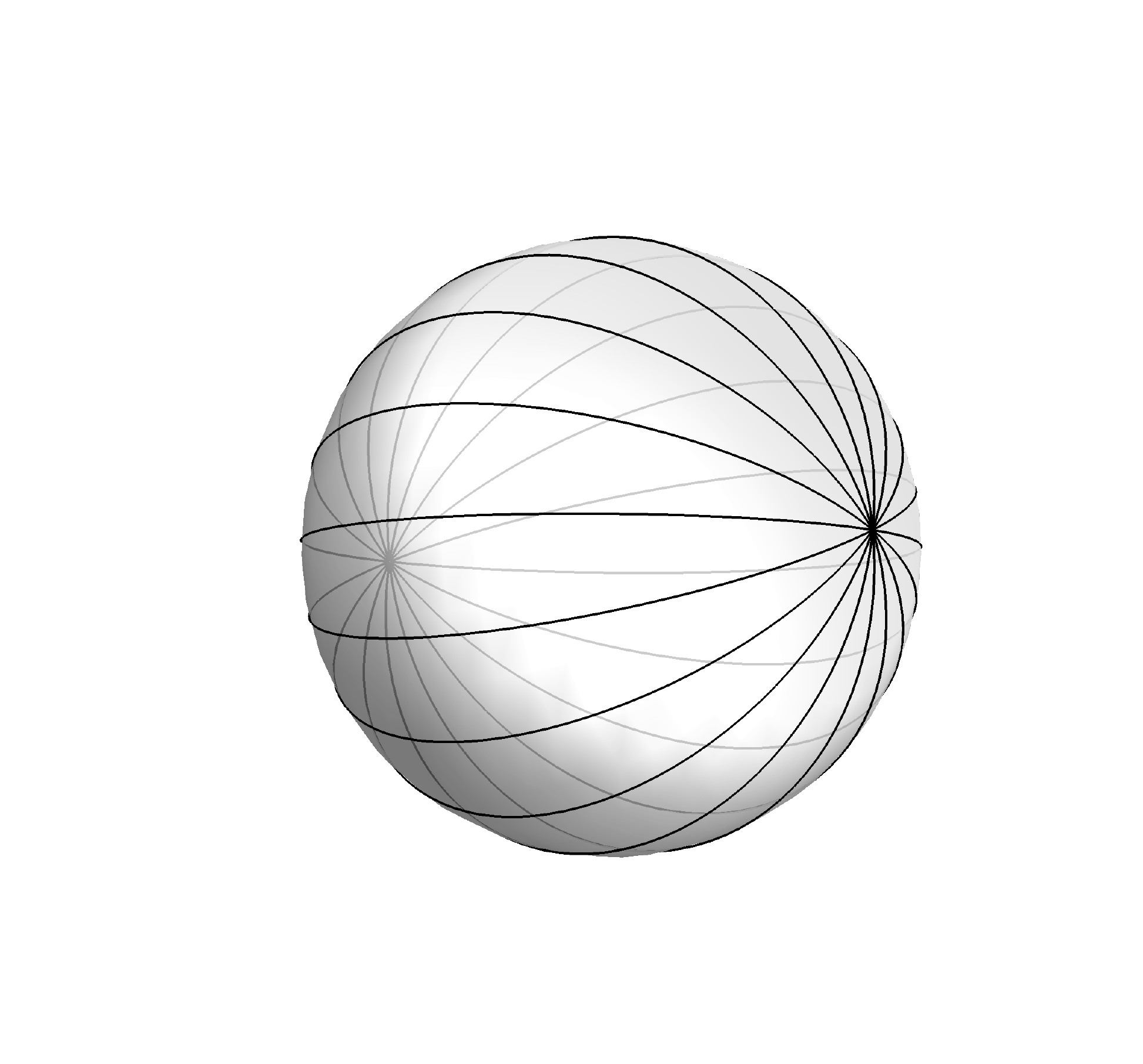}
 \end{center}
 \caption{ \footnotesize 
   Illustration of the topology of the  \emph{round}  Euclidean sphere ``path" $S^D$ for the case $D=2$.  
   The meridian arcs (half circles) each correspond to a patch of space at one time.
   They are topologically 1-balls, which all share the same 0-sphere boundary consisting of the two poles. Together they comprise a foliation of the 2-sphere that is degenerate at the poles, which constitute 
   the Euclidean horizon.
 }
 \label{fig2d} 
 \end{figure}

The paths in a path integral representation of $Z$
are periodic in time because the path integral is computing the trace.
At each time the configuration is 
 a Riemannian metric on 
a topological $(D-1)$-ball of space, whose
surface is a $(D-2)$-sphere.
In the saddle $D$-geometry the time translation becomes a Euclidean signature
rotation encircling the ball surface, 
which is a fixed point set of the rotation. 
The  manifold 
generated by rotating the ball through a complete 
time circle in the about the ball surface is topologically a $D$-sphere, $S^D$.  
This is easy to visualize in the 
$D=2$ case (see Fig. \ref{fig2d}),
where a complete rotation of a line segment about its endpoints sweeps out a topological 2-sphere.
The way it works for $D=3$ 
is explained in detail in section 5 of \cite{Banihashemi:2022jys}, which includes a possibly useful figure.   To see it in a general dimension $D$  one
can invoke the topological fact that the $(D-1)$-ball is   
the one point compactification of the half-space $\frac12\mathbb R^{(D-1)}$,
and rotating this through an extra dimension 
around its boundary yields the one point compactification
of $\mathbb R^{D}$, which is $S^D$.
 The boundaryless $D$-sphere topology for the paths 
in effect removes the need for a boundary condition at the surface of the spatial 
ball, and presumably corresponds to the condition that the spatial ball is 
smoothly embedded into the ambient space, like the dS static patch.

The partition function is dominated and thus well-approximated by the 
saddle---i.e., the solution to the vacuum Einstein equation---with minimal action. 
In the presence of a positive cosmological
constant $\Lambda$, 
the saddle with minimal action 
is a round $D$-sphere 
of radius $L=\sqrt{(D-1)(D-2)/(2\Lambda})$, a.k.a.\  Euclidean 
dS spacetime~\cite{Banihashemi:2022jys} \footnote{See also App.~A of the supplemental material for a proof, which includes Refs. \cite{akutagawa2019gap,Bishop,https://doi.org/10.48550/arxiv.1303.5390,Myers}.}. 
As mentioned above, the action of this
saddle is $-A/4 G$,  which (since there is no temperature dependence) implies that the 
entropy is given by $A/4 \hbar G$.
The cosmological constant (together with $\hbar G$) thus determines the dimension
of the Hilbert space of  all states of a topological ball of space in pure gravity.

In the limit $\Lambda\to0$, the de Sitter sphere saddle becomes a sphere of infinite radius
and the entropy is therefore infinite. 
We believe this result is correct, but it 
does not tell us what is the entropy
of a ball of a {\it fixed size}. 
Unlike the case in which
a particular   size is selected by the 
cosmological constant,   
any 
other size 
must be specified as an external constraint  that restricts the ensemble of states. 

Since  the entropy of a ball  is expected to be  proportional to 
the boundary area 
 one might 
think that the natural way to restrict 
the   size would be to restrict the boundary area.
A fixed   area constraint  can be imposed  using a Lagrange multiplier
term in the action, which contributes to the field
equations an effective energy-momentum tensor of a cosmic membrane
at the ball boundary.
The solutions to the resulting field equation have a conical 
defect at the membrane. 
It seems unlikely, however, that the presence of such a
conical defect allows for a
Euclidean solution to the $\Lambda=0$ vacuum Einstein equation on $S^D$. 
For instance, in $D=3$ dimensions such a metric would be locally flat everywhere except at the $S^1$ where there would be a conical defect,
but it can be shown that this is not possible~\cite{Hawkins}.    
The (presumed) absence of a saddle for the fixed area ensemble   suggests that the rigidity of this constraint entails quantum fluctuations that are too large to be compatible with a semiclassical description of the ensemble. According to the uncertainty relation, 
at fixed area one expects that the conjugate variable, which is the angle at which the maximal surface meets the edge of the causal diamond \cite{Carlip:1993sa},
is totally uncertain. 

One could instead fix the Euclidean spacetime volume, and this constrained partition function admits a saddle; however, 
this is not a physically valid way to implement the constraint that the ball has a particular ``size”. 
Since we are trying to count the dimension of the Hilbert space of the system, the size constraint should be imposed on the system itself, 
whose states are enumerated {\it at one time}.  
Moreover, the Euclidean spacetime geometry is relevant only in the saddle point approximation to the partition function being computed; 
the domain of integration in the path integral representation of the partition function is not 
Euclidean geometries \cite{Gibbons:1978ac,Schleich,Hartle-Schleich, Mazur-Mottola, Banihashemi:2022jys}. 
That said, the fixed spacetime volume partition function can perhaps be understood as a weighted sum dominated by spatial balls of a given size.
We are currently exploring this interpretation~\cite{JacVis2}. 

\noindent \emph{Partition function at fixed spatial volume.}
An apparently sensible way to fix the size of the system 
is to constrain its {\it spatial} volume.
The dimension of the Hilbert space of a quantum gravitational 
ball at fixed volume $V$ can be approximated as a 
path integral over metrics on the topological $D$-sphere that
admit a (degenerate) foliation by $(D-1)$-balls with
volume $V$ whose $(D-2)$-sphere boundaries all coincide.
 The paths are weighted by the exponential of minus the  
Einstein-Hilbert action with cosmological constant $\Lambda$ together with a Lagrange multiplier term implementing the volume constraint,
\begin{align}
\label{ZV}
Z[V,\Lambda]&=\int \! {\cal D}\l\, {\cal D}g\,\exp\Biggl[\frac{1}{16\pi \hbar G }\int d^D x\sqrt{g}(R-2\Lambda) \nonumber  \\
&+\frac{1}{\hbar}\int d \phi\, \lambda (\phi) \left ( \int \!d^{D-1} x\sqrt{\gamma}- V  \right) \Biggr] , 
\end{align}
where the contour of integration for the metric is assumed to pass through 
a Euclidean saddle. (See \cite{Banihashemi:2022jys} for a discussion
of the nature of the required contour deformation.) Here
$\phi$ is a periodic Euclidean coordinate, and   $\gamma_{ab} =  g_{ab} -N^2\phi_{,a}\phi_{,b}$  
is the induced metric on a constant $\phi$ slice, where 
$N:=(g^{ab}\phi_{,a} \phi_{,b})^{-1/2}$, so that $N d\phi$ is a unit 1-form. The integral of the Lagrange multiplier $\l(\phi)$
is  for each $\phi$  over a contour parallel to the imaginary axis, and thus introduces a Dirac delta function that imposes the constraint that the spatial volume of each $\phi$ slice is equal to $V$.   Following Gibbons and Hawking, we shall estimate 
$Z$ as in~\eqref{Zsim}, where $I_{\rm saddle}$ 
is the Einstein-Hilbert action evaluated at
the stationary point with the lowest action. 
The constraint term vanishes when the constraint is satisfied, so  
does not contribute to the saddle action. 

The saddle point equations are given by the volume constraint, together with the Euclidean 
Einstein equation sourced by a perfect fluid stress-energy tensor $T_{ab}$ with vanishing ``energy density,'' arising from the variation of the volume element in the volume constraint term,
\beq\label{EEq}
G_{ab}  + \Lambda g_{ab} = 8\pi G T_{ab} \,\,\,\, \text{with} \,\,\,\, T_{ab} = \frac{\lambda}{N} \gamma_{ab} =: P  \gamma_{ab}\,,
\eeq
where $P$ is the effective fluid pressure, 
and the $\lambda$ contour has been deformed so as to 
pass through the required  real value at the saddle. 
For each value of $V$, there is now a saddle, even if $\Lambda=0$. 
This is referred to as the method of constrained instantons~\cite{Affleck:1980mp,Cotler:2020lxj}. By analogy with the unconstrained $\Lambda>0$ case, where the saddle with the lowest action (i.e., the greatest volume) is the sphere, we presume that the dominating saddle is the most symmetrical one. We therefore look for a static, spherically symmetric solution with $\phi$ as a Killing coordinate, such that    $\lambda(\phi) =
\lambda=$ constant. Our metric Ansatz for the Euclidean saddle thus takes the form
\beq
ds^2 = N^2(r) d \phi^2 + h(r) dr^2 + r^2 d \Omega_{D-2}^2\,,
\eeq
where  $d\Omega_{D-2}^2$ is the line  element on a unit $(D-2)$-sphere, and we choose 
the period of the $\phi$ coordinate to be $\Delta\phi=2\pi$. 
With this symmetry ansatz, 
the equations to be solved are the Euclidean version of those 
of a static, spherically symmetric fluid star in $D$ spacetime dimensions,
but with different boundary conditions.

The spatial metric function $h(r)$ 
is determined by the $\phi\phi$ component of the Einstein 
equation, which receives no contribution from the 
fluid stress tensor since that
has vanishing ``energy density'' $T_{\phi\phi}$. 
The solution for $h(r)$ with $\Lambda >0$ is
thus the same as for $D$-dimensional dS space, 
\beq\label{h}
h(r) = (1 - r^2/L^2)^{-1},
\eeq
while that with $\Lambda=0$ corresponds to $L\to \infty$ and  that with $\Lambda <0$ to  $L\to iL$.
The equation for the metric function $N(r)$ is the (Euclidean, $D$-dimensional) Tolman-Oppenheimer-Volkoff (TOV) equation,
which   for vanishing energy density is identical to the corresponding Lorentzian equation. 
The boundary conditions for $N(r)$ arise from the
requirement that the ball boundary be a regular Euclidean horizon,
i.e., a fixed point set of the periodic $\phi$ translation symmetry where 
the metric is locally flat. 
This implies that the lapse $N(r)$ must vanish at some value  
$r=R_V$ (determined by the volume constraint) which defines the location of the horizon.
According to~\eqref{EEq} the pressure $P(r)$ and curvature 
must therefore diverge at the horizon, unless $\lambda=0$. 
Furthermore,    absence of a conical singularity at $R_V$ requires that the line element
in the $r$-$\phi$ subspace 
there takes the form of the Euclidean plane in polar coordinates,
$l^2 d\phi^2 + dl^2$,  where $l$ is the proper radial
distance from the Euclidean horizon. This fixes the coefficient of 
the linear term in a Taylor expansion of the lapse $N$ about $R_V$,
\beq\label{N'}
\frac{dN}{dl}\Big |_{l=0} = 1\,.
\eeq
We first solve the saddle point equations for $\Lambda=0$,   
and then generalize the solution to nonzero~$\Lambda$.

\noindent \emph{$\Lambda=0$ case.}
For $\Lambda=0$ \eqref{h} reduces to $h(r)=1$, so the spatial metric for the saddle is flat,
and the TOV equation with zero energy density is \cite{PoncedeLeon:2000pj}
 \beq
 \frac{dP}{dr} =- \frac{8 \pi G}{D-2}  P^2  r \,.
 \eeq
The general solution to this equation is 
 \beq \label{pressureflat}
 P (r)=- \frac{D-2}{4\pi G}\frac{1}{R_V^2-r^2}\, ,
 \eeq
  where $\RV^2 $ is an integration constant.
The condition that $P$  diverges somewhere  
requires that $R_V^2>0$, and the horizon is located at  $r=  \RV$. 
The value of $\RV $ is set by the volume constraint,
$\bigl(\Omega_{D-2}/(D-1)\bigr) \RV^{D-1}=V$, where $\Omega_{D-2}$ is the volume of a unit $(D-2)$-sphere. 
The lapse is given by $N(r) = \lambda/P(r)$, so the 
 condition \eqref{N'} implies that 
$\lambda = -1/(dP^{-1}/dl)_{l=0}$, i.e., 
\beq
\lambda = - \frac{1}{8\pi G  }\frac{D-2}{\RV}\,.
\eeq 
%
The 
Euclidean saddle metric is given by
 \beq \label{flatsaddle}
 ds^2 = \frac{1}{4R_V^2}( R_V^2-r^2)^2 d\phi^2 + dr^2 + r^2 d \Omega_{D-2}^2\,.
 \eeq
To estimate $Z[V]$ \eqref{ZV} we evaluate the on-shell 
 Einstein-Hilbert action with $\Lambda=0$,  
\beq \label{actionflat}
I_{\rm saddle} = - \frac{1}{16\pi G} \int d^D x\sqrt{g} R = \frac{D-1}{D-2}2\pi\lambda V
=-\frac{A_V}{4G}\,,
\eeq
where $A_V$ is the area of the $(D-2)$-sphere 
that forms the boundary of the spatial volume, 
i.e., the horizon of the Euclidean ``diamond''
with spatial volume $V$.
This action yields an approximation to the partition function,
in the zero-loop saddle-point approximation, 
\beq
Z[V]\approx\exp(A_V/4 \hbar G)\,.
\eeq%
The logarithm of the
dimension of the Hilbert space is thus given by the Bekenstein-Hawking 
entropy formula, with the horizon area equal to 
the boundary area of the saddle point ball metric.
This result for the action can also be seen from the fact that
the Euclidean Einstein-Hilbert action with Gibbons-Hawking-York boundary term, evaluated on a configuration that 
is independent of a periodic Euclidean ``time" coordinate,
has a Euclidean horizon of area $A$, 
and satisfies the vacuum Hamiltonian constraint, 
generally takes the form 
$I_{\text{saddle}}= \beta E_{\text{BY}} - A/4G$, with  $\beta$ the Euclidean time period, and $E_{\text{BY}}$ the Brown-York energy \cite{WhitingYork}.
This applies in the present case,   and   $E_{\text{BY}}$ vanishes as the boundary size goes to zero, so the action \eqref{actionflat} is reproduced in that limit.

 \begin{figure}[t!]
 \begin{center}
 \includegraphics[width=.5\textwidth]{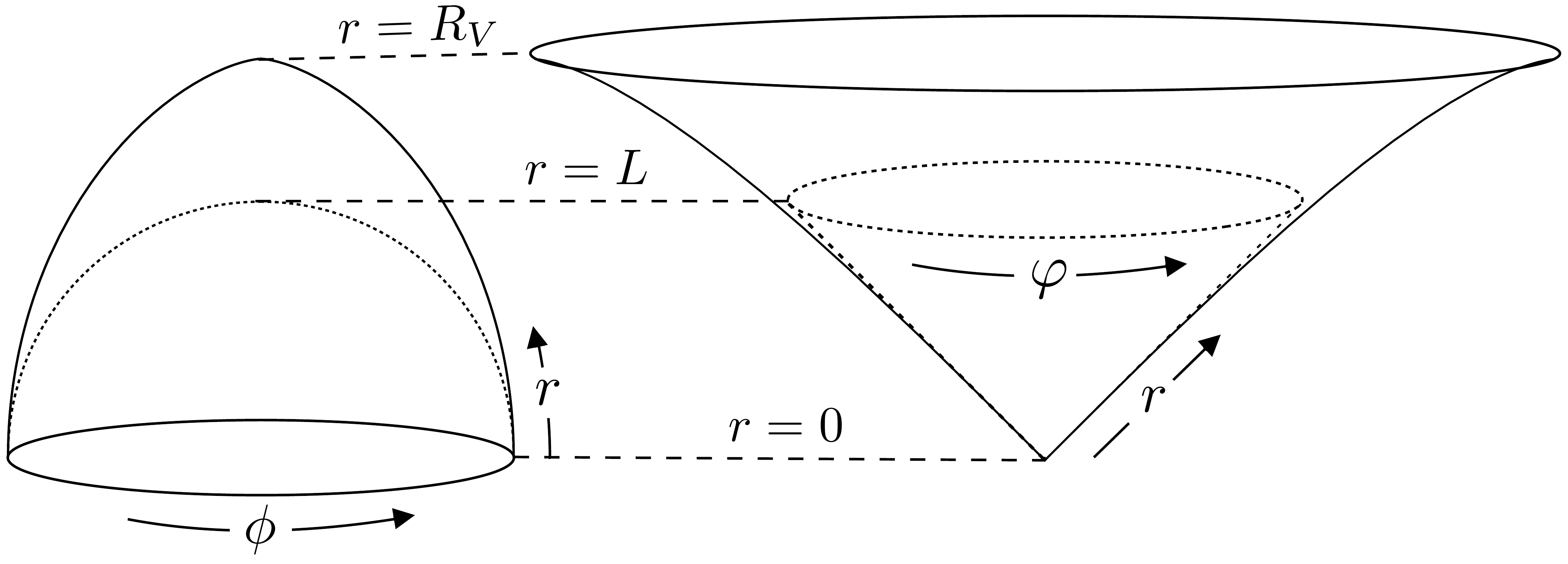} 
  \end{center}
 \caption{ \footnotesize  
  The Euclidean saddles for de Sitter with $\Lambda >0$ (dotted) and for fixed volume with $\Lambda=0$ (solid), with $\RV=2L$ so that the Euclidean time periods  match at the center of the saddle (the cap perimeter on the left). 
 The hemispherical caps are flat space embeddings of the intrinsic geometries of the time-radius ($\phi$-$r$) discs, with the horizons at the tips of the caps. At each point of the cap there is a round $(D-2)$-sphere, shown here on the right as    a circular section  of a cone 
  with azimuthal angle $\varphi$, whose radius 
  in the de Sitter case is equal to the vertical embedding height on the   hemispherical cap and
 in the $\Lambda=0$ case is equal to the   radial distance along the cap. 
  The fixed volume saddle has a mild curvature singularity at the horizon, 
 whereas the de Sitter saddle is   a round $D$-sphere which is everywhere smooth.  
 }
 \label{fig1} 
 \end{figure}

Let us discuss some  properties of the saddle. 
The saddle  has topology~$S^D$, is conformally flat \footnote{See App.~B in the supplemental material   for an explicit coordinate transformation to a conformally flat metric based on Ref. \cite{Casini:2011kv}.}, spatially flat, and spherically symmetric, and has a rotational Killing symmetry along the Euclidean time direction (see Fig.~\ref{fig1}).  
Further, the effective fluid pressure \eqref{pressureflat} is negative,
and diverges as the inverse proper distance to the edge of the ball (horizon), $P \sim -{1}/{(R_V - r)} $. Hence the energy-momentum tensor, and therefore the Ricci tensor, is singular on the horizon in the sense that it has eigenvalues that diverge as $1/ (R_V -r)$. 
However, this  curvature divergence  (which can be traced to 
a nonzero quadratic term in the Taylor expansion of the lapse function $N(r)$ 
about the horizon)
is sufficiently mild that the 
on-shell Einstein-Hilbert action is finite and given by \eqref{actionflat}.

Nevertheless, since the curvature diverges one should 
take into account higher derivative terms in the effective Lagrangian,
$L\sim R + \ell^2 R^2 + \dots$, with relative coefficients
determined by  some UV length scale $\ell$.
It is possible that such higher derivative terms 
allow for a regular saddle with everywhere finite curvature, but with divergent derivatives of curvature that match the divergence of the effective energy-momentum tensor.
If they do not, then our use of effective field theory appears 
inadequate to treat the problem. 
However, if they do, the entropy will be given by the Bekenstein-Hawking
term with the area of the horizon of the regular saddle, plus the higher curvature corrections to the horizon entropy
functional~\cite{Jacobson:1993xs,Wald:1993nt,Visser:1993nu}.
This computation would be under control in the
effective field theory provided that the higher curvature 
terms are systematically suppressed.
To estimate their contribution, 
note that the contribution of the $R^2$ term to the 
field equation behaves as $\sim \ell^2 \partial_r^2 R$, and  
the curvature of the uncorrected 
saddle diverges as $(R_V\rho)^{-1}$,
where $\rho := R_V-r$, so the contribution from the $R^2$ terms
scales as $\sim (\ell/\rho)^2 R$, which becomes 
comparable to the Einstein term when $\rho\lesssim \ell$.
At that point the curvature is~$\sim (R_V\ell)^{-1}$.
If the curvature saturates at this value,
the contribution of the $R^2$ term to the
entropy functional, which is 
the integral of $\sim \ell^2 R$ over the horizon,
would be of order $\ell/R_V$ relative to that of the 
Bekenstein-Hawking area term. 
Moreover, the contributions of
higher curvature terms would be 
suppressed by an additional factor of $\ell/R_V$ 
for each additional power of curvature in the Lagrangian, 
so that it would be consistent to truncate the effective 
field theory.

 
\noindent \emph{
$\Lambda \neq 0$ case.} 
For  nonzero $\Lambda$ a similar saddle exists, with action again given by $I_{\text{saddle}}=-A_V/4G$.
The most significant difference occurs for $\Lambda >0$, in which case 
the spatial geometry is a ball of volume $V$ embedded in a round $(D-1)$-sphere of radius $L$. 
$A_V$ grows with $V$ until $V$ reaches half the volume of that $(D-1)$-sphere, and then decreases
to zero as $V$ reaches the full $(D-1)$-sphere volume. (For details see the appendix.)

\noindent \emph{Discussion.}
To sum up, the quantum gravity partition function that counts the dimension of 
the Hilbert space of a spatial region with topology of a ball and fixed proper volume is approximated, in the leading order saddle point approximation,
by $\exp({A_V/4\hbar G})$, where $A_V$ is the area of the 
saddle ball boundary, up to higher curvature corrections suppressed by a UV length scale divided 
by the ball radius. 
For positive 
cosmological constant and
for volume $V> V_{\rm dS}$ (where $V_{\rm dS}$ is the volume of the
de Sitter static patch), 
$A_V$ and hence
the dimension of the Hilbert space {\it decreases} as $V$ increases.  The Hilbert space becomes 1-dimensional for $V=2V_{\rm dS}$, and no saddle exists for $V>2V_{\text{dS}}$. Hence, more space does
not always imply more states. 
The integral over all volumes, 
$\int dV \exp(A_V/4\hbar G)$, agrees, in the leading order 
saddle point approximation,
with the Gibbons-Hawking result \eqref{Zsim} for 
a ball of unconstrained volume, because the ensemble is dominated 
by ball geometries with surface area equal to that of the de Sitter horizon.

That the logarithm of the Hilbert space dimension in 
our nonperturbative framework matches the horizon 
entropy attributed to a semiclassical 
static patch in de Sitter space and other causal diamonds
supports the notion that the total dimension of the
Hilbert space is captured already at leading order by the 
exponential of the
semiclassical entropy. From this follows the surprising conclusion that 
the semiclassical, gravitationally dressed 
vacuum state of such a causal diamond 
is close to a maximally mixed state, a notion that has 
already been advanced for the case of a de Sitter static 
patch~\cite{Banks:2000fe,Banks:2003ta,Banks:2006rx,Fischler,Dong:2018cuv,Lin:2022nss,Banihashemi:2022jys,Chandrasekaran:2022cip}.

We have explicitly considered only the states of the gravitational field,
in a ball, but matter fields could be included without any modification
of the saddle calculation, provided the matter fields vanish in the saddle
configuration. The existence of extra states due to the matter
fields would be accounted for  at leading order  by the value of the low energy effective gravitational
constant $G$, which figures in the denominator of the Bekenstein-Hawking 
entropy~\cite{Susskind:1994sm,Jacobson:1994iw}. 
The entropy match discussed above 
thus lends nonperturbative quantum gravitational 
support to
the ``maximal vacuum entanglement hypothesis''~\cite{Jacobson:2015hqa}---that 
entanglement entropy of matter and gravity in small balls at fixed 
volume is maximized in the semiclassical, gravitationally dressed 
vacuum state. To elevate that support to full justification 
would require mastery of the corrections to the saddle point approximation.

\noindent \emph{Acknowledgments.}
We are grateful to Jan de Boer, Eli Hawkins, Harvey Reall, Erik Verlinde, Aron Wall, and Toby Wiseman for helpful discussions and/or comments. We also thank the anonymous referees for suggestions that led us to improve the presentation. The research of TJ is supported in part by National Science Foundation grant PHY-2012139. 
This work was done primarily while TJ was a Visiting Fellow Commoner at Trinity College, Cambridge. 
He is grateful to the college for hospitality and support.
This research was supported in part by Perimeter Institute for Theoretical Physics. 
Research at Perimeter Institute is supported by the Government of Canada through the Department of Innovation, Science and Economic Development and by the Province of Ontario through the Ministry of Research, Innovation and Science.
MRV is supported by  SNF Postdoc Mobility grant P500PT-206877 ``Semi-classical thermodynamics of black holes and the information paradox''.

\appendix

\setcounter{equation}{0}
\renewcommand{\theequation}{A\arabic{equation}}

 \section{Appendix: Derivation of the $\Lambda \neq 0$ saddle}
 
 In the presence of a positive cosmological constant, $h(r)$
 is given by \eqref{h}, so the spatial metric is that of a round $(D-1)$-sphere of radius $L$ and is  half covered by the range $r\in [0,L]$.  
The TOV equation for Einstein gravity with $\Lambda >0$ and zero energy density is \cite{PoncedeLeon:2000pj,deBoer:2009wk}
\beq\label{TOVL}
 \frac{dP}{dr} =- P \frac{\left(\frac{8 \pi GL^2}{D-2} P-1\right)r}{L^2-r^2}\,.
 \eeq
To uniquely label points on the $(D-1)$-sphere, we switch from $r$ to the polar angle coordinate $\chi\in[0,\pi]$ (with $r=L \sin \chi$), 
in terms of which the proper radial distance is $Ld\chi$. The case of $\Lambda<0$ is obtained by the replacement $(L,\chi)\rightarrow (iL,-i\chi)$ in the following formulae. For notational brevity we momentarily adopt units  with  $8\pi G /(D-2) =1$.
Then 
\eqref{TOVL} becomes
\beq\label{TOVL2}
 \frac{dP}{d\chi} =- P (L^2 P-1)\tan\chi\,,
 \eeq
and the solution is 
 \beq \label{pressure2}
P(\chi)=\frac{L^{-2}}{1- \cos\chi/\cos\chi_{{}_V}}, 
 \eeq
 where $\cos\chi_{{}_V}$ is an integration constant
 whose value is determined by the volume constraint, 
 \beq
 V=V(\chi_{{}_V}):=L^{D-1}\Omega_{D-2}\int_0^{\chi_{{}_V}} d\chi \, (\sin\chi)^{D-2}\,.
 \eeq 
The pressure is 
negative for $\chi_{{}_V} < \pi/2$, and positive for $\chi_{{}_V} > \pi/2$,  and  blows up as the reciprocal of the proper distance to the horizon at $\chi=\chi_{{}_V}$. 
The Lagrange multiplier $\lambda$ is fixed by the
condition \eqref{N'} at the horizon, 
 \beq \label{lambdadesitter}
 \lambda = -  L^{-1}\cot\chi_{{}_V}\,.
 \eeq
The Euclidean saddle metric is thus
 \beq \label{desittersaddlemetric}
  ds^2 =L^2 \left [   \left (\frac{\cos\chi -\cos\chi_{{}_V}}{\sin\chi_{{}_V}}\right)^2   d \phi^2 + d\chi^2  +\sin^2\!\chi\, d \Omega_{D-2}^2\, \right],
 \eeq
 with $\chi\in[0,\chi_{{}_V}]$.
This metric is also conformally flat, and it has a Killing horizon at $\chi=\chi_{{}_V}$. 
The saddle again has topology $S^D$, and a constant $\phi$ slice is a ball of
volume $V$ embedded in a round $(D-1)$-sphere of radius~$L$. In the limit $L \to \infty$ we have $\chi_{{}_V}\rightarrow0$, and holding $L\chi$ fixed \eqref{desittersaddlemetric} becomes the spatially flat metric \eqref{flatsaddle}.
For   $\chi_{{}_V}=\pi/2$ the metric becomes that of Euclidean de Sitter space (the round $D$-sphere of radius $L$) 
and $\lambda=0$, so the volume constraint plays no role in determining the saddle geometry. 
In the zero loop saddle point approximation, the partition function is thus unaffected 
by the constraint if the volume is set equal to that of (a maximal slice of) the dS static patch.

The Euclidean action of the saddle can be computed directly by using the saddle point equations~\eqref{EEq}, the pressure \eqref{pressure2}, and the metric \eqref{desittersaddlemetric}. In App.~C of the   supplemental material we present   an explicit computation of the on-shell action using the above coordinate system, and  in App.~D we derive the   on-shell action as well as a Smarr formula  and first law  using the Noether charge formalism \cite{Wald:1993nt,Iyer:1995kg} (see also \cite{Myers:1986un,Kastor:2009wy,Jacobson:2018ahi}). Alternatively, as explained below \eqref{actionflat}, we know from the 
general properties of the saddle that the action is given by $I_{\text{saddle}}=-A_V/4$. 
The horizon area of the saddle geometry is determined by  the parameters of the ensemble $V$ and $\Lambda$.
At fixed $V$ the area decreases as the cosmological constant increases. At fixed $\Lambda$, 
the area increases   with
volume until it reaches a maximum when $V=V(\chi_{{}_V} =\pi/2)$, such that the $(D-1)$-ball covers   half of 
a $(D-1)$-sphere of radius $L$, after which 
it decreases, reaching zero when $V=V(\chi_{{}_V}=\pi)$. 

In the case of $\Lambda<0$ the trig functions become hyperbolic trig functions, the saddle ball is embedded 
in hyperbolic space, and there is no upper limit to $\chi$ or to the size of the saddle ball as the volume grows.


\bibliography{diamondrefs}

 \newpage

\section{SUPPLEMENTAL MATERIAL}

\subsection{Appendix A: Round sphere  ($\Lambda>0$)  saddle has minimal action} 
\label{appA}

In $D=4$ dimensions the
only known saddle is the round sphere, but in $D=5$ to $D=9$ dimensions infinitely many saddles exist \cite{akutagawa2019gap}.
Nevertheless the round sphere saddle always has the minimal action, since it has the largest volume.
This can be seen as follows:
On a complete $D$-dimensional Riemannian manifold
on which each eigenvalue of the Ricci tensor is greater than or equal to  $(D-1)/L^2>0$,
the volume of any ball is always less than or equal to that of a ball of the same radius in the $D$-sphere with Ricci eigenvalues equal to  $(D-1)/L^2$ (this is the Bishop comparison theorem \cite{Bishop,https://doi.org/10.48550/arxiv.1303.5390}).
Moreover, in such manifolds any two points can be joined by a geodesic of length $\le\pi L$ (this is the Myers diameter bound \cite{Myers}), so a geodesic ball of radius $\pi L$ contains {\it all} points in the manifold. A ball of this radius just covers the $D$-sphere,
hence the total volume of the manifold is less than or equal to that of the sphere. The inequality is presumably never saturated. We thank Harish Seshadri for pointing us to this argument.

 \subsection{Appendix B: Euclidean saddle is conformally flat}
\label{sec:conformallyflat}

\setcounter{equation}{0}
\renewcommand{\theequation}{B\arabic{equation}}

In this appendix we show explicitly that the saddle geometry  for $\Lambda=0$, given by \eqref{flatsaddle} in the main body of the paper, is conformally flat. The metric of the saddle is
\beq
ds^2 =   \frac{R_V^2}{4} \left ( 1-\frac{r^2}{R_V^2}  \right)^2 d\phi^2 + dr^2 + r^2 d \Omega_{D-2}^2\, .
\eeq
In terms of the  radial coordinate  
\beq
r =  R \, \text{tanh}\left ( \frac{x}{2R_V}\right)\,,
\eeq
 the  line element  reads
\begin{align}
ds^2 =   &\frac{1}{4 \cosh^4 \!\left({x}/{2 R_V}\right)} \Big (   R^2_V d\phi^2 + dx^2 \nonumber \\
&+ R_V^2 \sinh^2 (x/R_V) d \Omega_{D-2}^2\Big)\,.
\end{align}
 The line element between brackets describes $S^1 \times H^{D-1}$, where $H^{D-1}$ is hyperbolic space, which is conformally flat, hence     the   saddle geometry above is also conformally flat. This can be  shown explicitly  by performing the coordinate transformation~\cite{Casini:2011kv}  
\beq
\tau = R_V \frac{\sin \phi}{\cosh (x/R_V) + \cos \phi}\,, \, y = R_V \frac{\sinh (x/R_V)}{\cosh (x/R_V) + \cos  \phi}\,,
\eeq
such that the   line element becomes a Weyl factor times the flat Euclidean metric   
\begin{align} 
\label{conformallyflat}
ds^2 &=\frac{R_V^2}{y^2} \left ( \frac{\sqrt{(y-R_V)^2+ \tau^2} - \sqrt{(y+R_V)^2+\tau^2}}{\sqrt{(y-R_V)^2+\tau^2}+\sqrt{(y+R_V)^2+\tau^2}}\right)^2 \times\nonumber\\&\times (d\tau^2 + dy^2 + y^2d \Omega_{D-2}^2) .
\end{align} 
The Killing vector field $\partial_\phi$  of the saddle in   these coordinates is
\beq 
\zeta = \frac{1}{2 R_V} \left [ (R_V^2- y^2 +\tau^2)\partial_\tau + 2\tau  y  \partial_y \right].
\eeq
 This is  a  conformal Killing vector of the flat metric $d\tau^2 + dy^2 + y^2d \Omega_{D-2}^2$, though it is a Killing vector of   the entire metric \eqref{conformallyflat}.
Under $\tau\rightarrow -i\tau$, it becomes ($i$ times) the conformal Killing vector 
of the Minkowski metric $-d\tau^2 + dy^2 + y^2d \Omega_{D-2}^2$ 
that preserves the causal diamond of radius $R_V$, centered at $\tau=y=0$.

\section{Appendix C: On-shell Euclidean action of the $\Lambda>0$ canonical ensemble saddle}
\label{onshellaction}

\setcounter{equation}{0}
\renewcommand{\theequation}{C\arabic{equation}}

In the main text we evaluated the on-shell Euclidean action of the $\Lambda>0$ saddle by invoking a general result showing that it has the form $\beta E_{\text{BY}} - A/4G$. In this appendix we verify this result by explicit evaluation of the action using a particular static coordinate system. 

The off-shell Euclidean action is given by the sum of the Einstein-Hilbert action with cosmological constant and the Gibbons-Hawking-York boundary term  at the York boundary where the canonical ensemble parameters are specified, 
\begin{align}
I=I_{\text{EH}}  + I_{\text{GHY}}= &- \frac{1}{16 \pi G } \int_{\mathcal M_\text{E}} d^D x \sqrt{g} (R - 2 \Lambda) \nonumber \\
&- \frac{1}{8\pi G} \int_B d^{D-1} x \sqrt{\gamma} K_B \,.
\end{align}
Here $\mathcal M_{\text E}$ is the Euclidean spacetime region with boundary $B$, and  
$K$ is the trace of the extrinsic curvature of the boundary $B$ as embedded in $\mathcal M_{\text E}$,  defined with respect to the normal pointing outward from the system. 

We are interested here in the action of the saddle discussed in the main text.
We first evaluate the action on shell  with a  
 finite sized boundary  (for which the saddle topology is $D^2\times S^{D-2}$)
 and later take the limit of vanishing boundary size (for which the  saddle  topology is $S^D$). 
From the saddle point equations 
\eqref{EEq}   
it follows that the bulk action on shell is given by 
 \beq
I_{\text{EH}} =     \int_{\mathcal M_\text{E}} d^D x \sqrt{g} \left ( \frac{D-1}{D-2} P - \frac{\Lambda}{(D-2)4 \pi G} \right)\,.
\eeq
Inserting the pressure \eqref{pressure2} 
 (with the factor     $8\pi G/(D-2)$ restored using dimensional analysis)  
and the  line element \eqref{desittersaddlemetric}
for the $\Lambda>0$ saddle,  yields  
\begin{align}
 \label{bulkaction}
I_{\text{EH}} &= - \frac{L^{(D-2)}\Omega_{D-2}}{4G} \frac{D-1}{\sin\chi_{{}_V}} 
\int_{\chi_{{}_B}}^{\chi_{{}_V}}d\chi\, \cos\chi\,(\sin\chi)^{D-2} \nonumber \\
&=\frac{1}{4G}\left(\frac{\sin\chi_{{}_B}}{\sin\chi_{{}_V}}A_B - A_V\right)
\,, 
\end{align}
where $A_{B}$ and $A_{V}$ are the areas of the York boundary cross section and saddle sphere surface, respectively. 
Further, for a metric of the form of the Euclidean saddle metric \eqref{desittersaddlemetric}
the trace $K_B$ of the $(D-1)$ dimensional extrinsic curvature of the boundary is 
given by 
\beq
K_B = -\frac{d(NA_{D-2})/Ld\chi}{NA_{D-2}}= k_B -\frac{dN/d\chi}{NL},
\eeq
where $k_B$ is the trace of the $(D-2)$ dimensional extrinsic curvature of a constant $\phi$ slice of the boundary.
Hence the boundary action is given by
\begin{align}
I_{\text{GHY}} &=-\frac{1}{8\pi G} \int_B d^{D-1} x \sqrt{\gamma} \left(k_B -\frac{dN/d\chi}{NL}\right)\nonumber\\
&= \beta E_{\rm BY} - \frac{1}{4G}\frac{\sin\chi_{{}_B}}{\sin\chi_{{}_V}}A_B\,,\label{bdyaction}
\end{align} 
where $\beta$ is the Euclidean time period and $E_{\text{BY}}$ is the Brown-York energy, 
\beq
\beta = 2 \pi N_B\,, \qquad E_{\text{BY}} = \frac{D-2}{8\pi GL} \frac{\cos\chi_{{}_B}}{\sin\chi_{{}_B}}A_B\,.
\eeq
Adding \eqref{bulkaction} and \eqref{bdyaction}   we find that the total on-shell Euclidean action becomes
\beq
I_{\text{saddle}}=\beta E_{\text{BY}}  - \frac{A_V}{4G}\,.
\eeq
In the limit that the boundary radius $L\sin\chi_{{}_B}$ goes to zero,  
the Brown-York energy vanishes if $D\ge 4$ and the action reduces to $-A_V/4G.$

\section{Appendix D: On-shell  action, Smarr relation, and first law via Noether charge}
\label{appsmarr}

\setcounter{equation}{0}
\renewcommand{\theequation}{D\arabic{equation}}

In this Appendix we derive the action, the Smarr relation, and the first law for the Euclidean saddle 
using the Noether charge formalism \cite{Wald:1993nt}.
The diffeomorphism Noether current $(D-1)$ form $j_\xi$  associated to a general  vector field $\xi$  is defined as
\beq \label{noethercurrent}
j_\xi := \theta (\psi,\mathcal L_\xi \psi) - \xi \cdot L \,,
\eeq
with $\psi$ the dynamical fields,  $\theta$ the symplectic potential $(D-1)$-form and $L$ the Lagrangian $D$-form, whose field variation is $\delta L = E \delta \psi + d \theta (\psi, \delta \psi)$. 
In our case the dynamical fields are the metric $g_{ab}$ and the Lagrange multiplier $\lambda$, and the  Lagrangian $D$-form consists of the Einstein-Hilbert term plus a cosmological constant and a Lagrange multiplier term
imposing the constraint that the spatial volume   of each  constant $\phi$ surface  is equal to $V$,
\beq\label{L}
L = L_{\text{EH}} + L_\lambda= -\epsilon\,(R- 2\Lambda)/{16\pi G}+ \lambda (\phi) d \phi \wedge(\epsilon_\Sigma - v_\Sigma)\,,
\eeq
where $\epsilon$ is the volume form, 
$\epsilon_\Sigma:=|\partial_\phi|^{-1}\partial_\phi\cdot\epsilon$ and 
$v_\Sigma := V \epsilon_\Sigma/\int_\Sigma \epsilon_\Sigma$. The Noether current associated to the full Lagrangian is not closed on shell, i.e., $dj_\xi = - E \mathcal L_\xi \psi + \delta_\xi L - \mathcal L_\xi L \neq 0 $, because of the non-locality of the spatial volume $v_\Sigma$. The obstruction is that the    variation of the Lagrangian $\delta_\xi L$ produced by the vector field induced variation  of the fields $\delta_\xi \psi = \mathcal L_\xi \psi$ is not equal to the Lie derivative of the Lagrangian with respect to the  vector field, $ \delta_\xi L \neq \mathcal L_\xi L$ \footnote{For example,  $\mathcal L_\xi L = d(\xi\cdot L)$ vanishes at a point where $\xi$  and its derivative vanish, but $\delta_\xi L$  would not vanish at that point since $L$  depends nonlocally on the geometry.}. Nevertheless, the current associated to the Einstein-Hilbert term satisfies the off-shell identity \cite{Iyer:1995kg}
\beq \label{offshellnoether}
j_{\xi}^\text{EH} = dQ_{\xi}^\text{EH} +2 \xi^a E_{ab}^\text{EH}\epsilon^b\,,
\eeq
where indices on $\epsilon$ refer to some of the $D$-form indices, the rest being implicit, 
$Q_\xi^{\text{EH}}=\frac{1}{16\pi G}\epsilon_{ab}\nabla^a\xi^b$ is the   
Noether charge $(D-2)$-form, and   $E_{ab}^{\text{EH}}= \frac{1}{16\pi G} G_{ab}  $ is the metric variational derivative of the Einstein-Hilbert term.  If we take $\xi$ to be the Killing vector $\partial_\phi$ 
of the Euclidean saddle, and use the Einstein equation, then the second term on the right-hand side of \eqref{offshellnoether} vanishes since the effective energy-momentum tensor \eqref{EEq} 
has only spatial components. 

The on-shell Euclidean action for a $\phi$-independent solution is given by
\beq\label{I}
I = \int L = \int d\phi\wedge (\partial_\phi\cdot L) =  -\frac{2\pi}{\kappa} \int_\Sigma j_{\partial_\phi} \,,
\eeq
where $\kappa$ is the horizon surface gravity with respect to the Killing vector
$\partial_\phi$ (which we set to unity in the main text), and in the last step we used 
\eqref{noethercurrent} and the fact that $\theta$ depends linearly on 
it second argument which vanishes since $\partial_\phi$ is a Killing vector.
Using the on-shell version of the Noether identity \eqref{offshellnoether}
it follows that the action of the saddle is proportional to the horizon integral of the Noether charge,
\beq
I_{\text{saddle}} =  - \frac{2\pi}{\kappa} \oint_{\partial \Sigma} Q_\xi\,.
\eeq
Furthermore, the integral of the Noether charge over the ball boundary is 
\beq \label{noethercharge}
\oint_{\partial \Sigma} Q_\xi^{\text{EH}} = \frac{\kappa A_V}{8\pi G}\,,
\eeq
so the saddle action is minus the Bekenstein-Hawking entropy,
\beq\label{Isaddle}
I_{\text{saddle}} =   - \frac{A_V}{4 G}\,.
\eeq

The Smarr relation follows from integrating the on-shell   identity $ j_\xi^{\text{EH}}=dQ_\xi^{\text{EH}}$,
with $\xi = \partial_\phi$, over  a spatial $(D-1)$-dimensional ball $\Sigma$ with boundary $\partial \Sigma$
\beq \label{noetheridentity}
\int_\Sigma j_\xi^{\text{EH}} =\oint_{\partial \Sigma} Q_\xi^{\text{EH}} \,. 
\eeq
The right hand side is given by \eqref{noethercharge}.
As for the left hand side, 
since  $\xi$ is a Killing vector it follows that 
\beq
j_\xi^{\text{EH}} = -\xi\cdot L_{\text{EH}} = \xi\cdot\epsilon(R-2\Lambda)/16\pi G\,.
\eeq
The metric field equation \eqref{EEq} implies  
\beq
\frac{1}{16\pi G} (R-2\Lambda)=  - \frac{\lambda}{N} \frac{D-1}{D-2} + \frac{\Lambda}{4 \pi G (D-2)} \,,  
\eeq
with $N=\sqrt{g_{ab}\xi^a \xi^b}$. 
The integral of the Noether current is hence 
\beq \label{noethercurrentsaddle}
\int_\Sigma j_\xi^{\text{EH}} =  -\frac{D-1}{D-2} \lambda V + \frac{\Lambda V_\xi  }{4 \pi G (D-2)}\,.
\eeq
Here, $V_\xi$ is the Euclidean   ``Killing volume'' defined as \cite{Jacobson:2018ahi}
\beq
V_\xi \equiv \int_\Sigma \xi \cdot \epsilon \,.
\eeq
The Lagrange multiplier $\lambda$ of   the saddle is related to the surface gravity $\kappa$ and the trace 
$k$ 
of the extrinsic curvature of the boundary $\partial \Sigma$    embedded in the $(D-1)$-ball $\Sigma$, with respect to the normal pointing away from $\Sigma$,
by 
\beq
\lambda = - \frac{\kappa k }{8\pi G}\,. 
\eeq
This follows by writing the previously found solution for $\lambda$ \eqref{lambdadesitter} in a covariant way, but it can also be derived directly from the action: On a $\phi$-independent configuration, the derivative of the action $\int L$ with respect to the fixed volume $V$,  with the Lagrangian $L$ given by 
\eqref{L},  is $(2\pi/\kappa)\lambda$. On the other hand the on-shell action is given by \eqref{Isaddle}, so it
follows that on shell we have $(2\pi/\kappa) \lambda = -(1/4G)dA_V/dV = -k/4G$. 
The integral   identity \eqref{noetheridentity} thus yields the Smarr relation 
\beq
(D-1) \kappa k V   +2 \Lambda V_\xi   =(D-2)  \kappa A_V \,.
\vspace{2mm}
\eeq
This agrees precisely with the Smarr relation that we previously found for the maximal slice of causal diamonds in maximally symmetric spacetimes \cite{Jacobson:2018ahi} (the reason being that the conformal Killing symmetry
of a causal diamond becomes  an ``instantaneous'' Killing symmetry at the maximal slice).  
The first law can be derived from the Smarr relation by applying Euler's theorem 
to the function $A(V,\Lambda)$ (cf.\  \cite{Myers:1986un,Kastor:2009wy,Jacobson:2018ahi}), which yields
\beq
\kappa \delta A_V = \kappa k \delta V - V_\xi \delta \Lambda\,.
\eeq

\end{document}